\documentclass[reqno,12pt]{amsart}
\usepackage{fullpage}
\usepackage{amsmath,amsfonts,amssymb}

\numberwithin{equation}{section}

\newtheorem{theorem}{Theorem}[section]{\bf}{\it}

{\bf}{\it}

\newtheorem{corollary}[theorem]{Corollary}{\bf}{\it}

\newtheorem{proposition}[theorem]{Proposition}{\bf}{\it}

\def\Ref#1{Ref.~\cite{#1}}

\def\eos/{equation of state}
\def\esos/{equations of state}
\def\com/{constant of motion}
\def\csom/{constants of motion}

\def\Rnum{{\mathbb R}}
\def\d#1{{\mathbf d}{#1}}
\def\lieder#1{{\mathcal L}_{#1}}
\def\hook{\rfloor}

\def\Esp{{\mathcal E}}

\def\Rop{{\mathcal R}}
\def\I{{\mathcal I}}

\def\vol{{\rm vol}}
\def\J{{\mathrm J}}
\def\rvec{\hat{r}}
\def\k{\kappa}
\def\M{{\mathcal M}}
\def\A{{\mathcal A}}

\begin{document}

\title{New conserved integrals and invariants\\ of radial compressible flow\\ in $n>1$ dimensions}
\author{
Stephen C. Anco${}^1$
\lowercase{\scshape{and}} 
Sara Seifi${}^1$
\lowercase{\scshape{and}} 
Amanullah Dar${}^{1,2}$
\\\lowercase{\scshape{
${}^1$Department of Mathematics and Statistics, Brock University, 
St. Catharines, ON Canada}}\\
\\\lowercase{\scshape{
${}^2$Department of Mathematics, University of Kotli, 
Azad Jammu and Kashmir, Pakistan}}
}

\begin{abstract}
Conserved integrals and invariants (advected scalars) are studied 
for the equations of radial compressible fluid/gas flow in $n>1$ dimensions. 
Apart from entropy, which is a well-know invariant, 
three additional invariants are found from an explicit determination of 
invariants up to first-order.
One holds for a general \eos/, and the two others hold only for entropic \esos/. 
A recursion operator on invariants is presented, 
which produces two hierarchies of higher-order invariants. 
Each invariant yields a corresponding integral invariant, 
describing an advected conserved integral on transported radial domains. 
In addition, a direct determination of kinematic conserved densities 
uncovers two ``hidden'' non-advected conserved integrals: 
one describes enthalpy-flux, holding for barotropic \esos/;
the other describes entropy-weighted energy, holding for entropic \esos/. 
A further explicit determination of a class of first-order conserved densities 
shows that the corresponding non-kinematic conserved integrals 
on transported radial domains 
are equivalent to integral invariants, modulo trivial densities. 
\end{abstract}

\maketitle

\section{Introduction}

A significant interest in conserved integrals and invariants of 
inviscid compressible fluid/gas dynamics in $n>1$ dimensions
has been on-going for several decades, 
\cite{Dez,Ser,KheChe,God,Kur,AncDar2009,AncDar2010,Anc,AncDarTuf,BesFri}, 
motivated by Hamiltonian structures and group-theoretical properties of fluid mechanics \cite{Arn1966,Arn1969,HolKup83c,Ver,Kup,ArnKhe}. 
Conserved integrals are global balance equations (conservation laws) of the form
$\frac{d}{dt} \int_{V(t)} \mathrm{C}\,d^nx = -\oint_{\partial V(t)} \vec{\mathrm{F}}\cdot\hat n\,d^{n-1}A$
on any domain $V(t)$ transported in the flow, 
where $\mathrm{C}$ is the conserved density 
and $\vec{\mathrm{F}}$ is outward flux on the moving boundary $\partial V(t)$. 
These integrals describe physically conserved quantities. 
Invariants are quantities (scalars, vectors, 1-forms, etc.) $\J$ that are advected by the flow, 
$\partial_t \J + \lieder{\vec{u}} \J=0$,
where $\lieder{\vec{u}}$ is the Lie derivative with respect to the velocity $\vec{u}$. 
These quantities are, physically, frozen into the flow. 
There is a direct correspondence between scalar invariants 
and conserved integrals whose moving flux vanishes, called integral invariants, 
as given by $\mathrm{C}= \rho \J$ in terms of the density $\rho$, with $\vec{\mathrm{F}}\equiv 0$. 

Currently, a complete description is known for 
conserved integrals of kinematic type and vorticity type \cite{AncDar2009,AncDar2010}, 
and a geometric procedure is known for generating higher-order invariants 
\cite{TurYan,BesFri,AncWeb2020} from the set of basic invariants. 
Yet some open questions remain to be addressed: 
What is the complete set of all invariants up to first-order? 
Does every non-trivial scalar invariant yield a non-trivial integral invariant? 
Are there any conserved integrals with conserved densities of low order 
other than the kinematic and vorticity ones? 
Do conserved integrals exist having higher-order conserved densities? 

The present paper fully answers these questions for $n$-dimensional 
compressible radial fluid/gas dynamics. 
There are several interesting reasons to study radial flows. 
A priori, since they arise from radial reduction of the $n$-dimensional governing equations, 
additional conserved integrals and invariants may exist which are not inherited through this reduction. 
The resulting radial equations have physical applications to 
explosive and implosive flows (see e.g. \cite{Tay,Kel,Wel,JenTsi,MerRapRodSze,Whi-book,Can-book}), 
as well as numerous engineering applications such as 
radial flow turbines, pumps, impellers, and compressors
(see e.g. \cite{Aun-book,CasRob-book}). 
Moreover, the radial equations describe the simplest case of more general zero-vorticity flows. 

For generality, radial flows with a general \eos/ will be considered. 
In the case of fluids, the pressure will be a general function of entropy (or temperature) in addition to density; 
likewise in the case of gas dynamics, the sound speed will be a general function of pressure and density (or temperature). 
This will encompass all of the familiar \esos/ ---
barotropic, polytropic, ideal gas --- as well as the case of an entropic \eos/ in which 
the pressure is a function only of entropy. 
Specifically, existence of conserved integrals and invariants that may arise for any special \esos/ will be determined. 

Several new results are obtained for invariants and conserved integrals. 

Firstly, 
all local invariants up to first-order are derived, 
and a recursion operator is found which generates higher-order invariants. 
This operator is shown to yield all second-order invariants. 
The set of basic invariants is found to include a new first-order invariant 
which exists for a general \eos/,
and two other new first-order invariants which exist in the case of an entropic \eos/. 
None of these invariants are inherited from $n$-dimension (non-radial) flow. 

Secondly, 
the new invariants give rise to corresponding integral invariants. 
By use of the recursion operator, 
an infinite hierarchy of integral invariants is obtained for a general \eos/,
and two additional infinite hierarchies are obtained for an entropic \eos/.
All of these integral invariants represent new, advected quantities. 

Thirdly, 
conserved integrals with local densities of at most first-order are derived 
using an extension of the method developed for kinematic densities 
in \cite{AncDar2009,AncDar2010}. 
In addition to kinematic ones inherited from non-radial flow, 
this yields two new conserved integrals, 
one describing enthalpy-flux in the case of a barotropic \eos/,
and the other describing entropy-weighted energy in the case of an entropic \eos/. 
Moreover, the only non-kinematic (first-order) conserved integrals which arise 
are shown to be equivalent to integral invariants, modulo trivial densities. 

An example of a new conserved integral with a nonlocal conserved density, 
and a new nonlocal invariant, are shown in remarks at the end. 

The paper is organized as follows. 
In Section~\ref{sec:eqns}, the radial reduction of the governing equations is summarized, 
and the radial formulation of conservation laws and conserved integrals is explained. 
In Section~\ref{sec:invs}, the radial formulation of scalar, vector, and 1-form invariants is presented, followed by the results of a classification of local invariants up to first-order.
The recursion operator is constructed from a novel relationship between scalar and vector invariants. 
In Section~\ref{sec:conslaws}, the results on conserved integrals are presented. 
Invariant integrals are summarized in \ref{sec:invintegrals}, 
while kinematic and first-order conserved integrals are classified in \ref{sec:noninvintegrals}. 
Scaling properties of these conserved integrals are discussed in \ref{sec:scaling}, 
and two applications of the lowest-order new conserved integrals 
are outlined for the analysis of smooth flows, shocks, and self-similar flows. 

In Section~\ref{sec:remarks}, some concluding remarks are made. 
Two appendices provide remarks on computational aspects of the main results, 
which have been carried out by use of Maple.

\section{Governing equations and preliminaries}\label{sec:eqns}

For fluid flow in $n$ dimensions without boundaries, 
the dynamical variables are functions of position $\vec{x}$ in $\Rnum^n$ and time $t$:
velocity $\vec{u}(\vec{x},t)$;
density $\rho(\vec{x},t)$;
pressure $p(\vec{x},t)$. 
Attention is restricted to locally adiabatic flows, with specific entropy $S(\vec{x},t)$. 

The governing equations for compressible flows are given by the Euler equations
\begin{gather}
\vec{u}_t + \vec{u}\cdot\nabla \vec{u} = -\frac{1}{\rho}\nabla p ,
\label{velocity.eqn}
\\
\rho_t + \nabla\cdot(\rho \vec{u}) = 0 ,
\label{density.eqn}
\\
S_t + \vec{u}\cdot \nabla S = 0 .
\label{entropy.eqn}
\end{gather}
This system is closed by specifying an \eos/, which in general is given by 
\begin{equation}
p=p(\rho,S) .
\label{eos}
\end{equation}

A worthwhile remark is that all thermodynamic quantities can be obtained 
in terms of the internal energy $e(\rho,S)$ 
through the thermodynamic relation 
\begin{equation}\label{thermo.rel}
T\,dS = de + p\,d(1/\rho)
\end{equation}
where $T$ is the local temperature. 
In particular, from the \eos/, 
the internal energy is given by 
\begin{equation}
e(\rho,S) =\int (p/\rho^2)\,d\rho
\end{equation}
which determines the temperature
\begin{equation}
T(\rho,S) =\frac{\partial e}{\partial S}\Big|_{\rho} = \int (p_S/\rho^2)\,d\rho . 
\end{equation}

As is well known \cite{Whi-book},  
the Euler equations for compressible fluid flow are equivalent to the equations of gas dynamics 
as given by the velocity equation \eqref{velocity.eqn}, the density equation \eqref{density.eqn}, and the dynamical equation for pressure
\begin{equation}
p_t + \vec{u}\cdot\nabla p + a^2 \rho \nabla\cdot\vec{u}  = 0 
\label{pressure.eqn}
\end{equation}
where 
\begin{equation}
a=a(\rho,p) >0
\label{soundspeed}
\end{equation}
is the sound speed. 
The pressure equation can be derived from the \eos/ \eqref{eos}
by use of the implicit function theorem to obtain $S=F(\rho,p)$, 
which is then substituted into the entropy equation \eqref{entropy.eqn}
and simplified using the density equation \eqref{density.eqn}, 
with 
\begin{equation}
a^2=-F_\rho/F_p = \frac{\partial p}{\partial\rho}\Big|_{S=F(\rho,p)} . 
\label{asq.rel}
\end{equation}
Conversely, the entropy equation can be recovered from the pressure equation \eqref{pressure.eqn} 
by solving $F_\rho + a^2(\rho,p) F_p=0$ to obtain $S=F(\rho,p)$, 
which is then observed to satisfy the entropy equation 
as a consequence of the density and pressure equations. 

It will be useful to summarize the most common \esos/ \eqref{eos} 
arising in physical applications for the Euler fluid equations 
and their counterparts for the sound speed \eqref{asq.rel} in gas dynamics. 
Hereafter,
\begin{equation}\label{materialder}
\frac{d}{dt}=\partial_t + \vec{u}\cdot\nabla
\end{equation}
denotes the material derivative. 

(A) non-barotropic \eos/: 
$p=P(\rho,S)$, $P_S\not\equiv0$; 
$a^2=P_\rho(\rho,F(\rho,p))$, $F_p\not\equiv0$.
Physically, this describes non-isothermal flows, namely
$\frac{d}{dt}T\not\equiv0$. 

(B) barotropic \eos/:
$p=P(\rho)$, $P'\not\equiv0$;
$a^2=P'(\rho)$. 
These flows are isothermal on streamlines, namely 
$\frac{d}{dt}T=0$. 

(C) polytropic (power law) \eos/:
$p=\k(S) \rho^{1+\gamma}$, $\k(S)\not\equiv0$ and $\gamma=$const.; 
$a^2=(1+\gamma)p/\rho$. 
In these flows, 
$\frac{d}{dt}T=-\gamma T \nabla \cdot\vec{u}$. 

(D) ideal (perfect) gas law \eos/:
$p=k\rho T$, 
where $k$ is Boltzmann's constant. 
This is a special case of a polytropic flow 
in which $\gamma=\frac{2}{n}$ and $\k(S)= \exp(\frac{\gamma}{k}S)$. 
It is also sometimes called the ideal polytropic case. 

(E) entropic \eos/:
$p=\k(S)$, $\k'(S)\not\equiv 0$; 
$a=0$. 
In these flows, sound does not propagate, which describes a shockless gas,
with $\frac{d}{dt}T=T \nabla \cdot\vec{u}$. 
Physical applications arise, for example, 
in modelling cavitation processes, 
behaviour near liquid-gas critical point transitions, 
some models of solar wind, 
and cold dark matter models in cosmology. 

An important consequence of the equivalence between 
the respective governing equations of compressible fluid flow and gas dynamics 
is that every conservation law and invariant admitted by the fluid equations
holds for the gas dynamics equations. 
This correspondence can be made explicitly manifest
when a conservation law or an invariant is expressed 
solely in terms of $t$, $r$, $\vec{u}$, $\rho$, $p$, $e$.

\subsection{Radial flow}

The reduction of the governing equations \eqref{velocity.eqn}--\eqref{entropy.eqn}
to radial flows consists of taking $\vec{u}$, $\rho$, $S$ to be functions of only $r=|\vec{x}|$ and $t$, 
and requiring $\vec{u}$ to be parallel to the radial unit vector $\rvec=(1/r)\vec{x}$. 
Thus, 
\begin{equation}
\vec{u}=\vec{u}(r,t) = U(r,t) \rvec,
\quad
\rho=\rho(r,t),
\quad
S=S(r,t) . 
\label{radial.reduc}
\end{equation}  
These variables satisfy the reduced Euler equations
\begin{align}
& 
U_t+ U U_r + ( p_S S_r + p_{\rho} \rho_r )/\rho =0, 
\label{radialU.eqn}
\\
& 
\rho_t + (U\rho)_r +\tfrac{n-1}{r} U \rho = 0, 
\label{radialrho.eqn}
\\
& 
S_t + U S_r = 0. 
\label{radialS.eqn}
\end{align}

Note that the vorticity of a radial flow is zero, 
since 
$\nabla\wedge\vec{u} = -\rvec\wedge\nabla U= -(\rvec\wedge\rvec) U_r= 0$,
due to $\nabla\wedge \rvec=0$ and $\nabla r = \rvec$. 
Zero vorticity flows have the general form $\vec{u}=\nabla \phi$ 
for any function $\phi(\vec{x},t)$, with $\phi=\phi(r,t)$ in the particular case of radial flows.

Radial reduction \eqref{radial.reduc} coincides with spherical symmetry reduction when $n>2$
but it is more restrictive when $n=2$. 
Spherical symmetry is defined by invariance of the fluid variables under the rotation group $SO(n)$: 
$\lieder{\vec{\xi}}\vec{u}=0$, $\lieder{\vec{\xi}}\rho=0$, $\lieder{\vec{\xi}}S=0$, 
for all vector fields $\vec{\xi}$ that generate rotations in $\Rnum^n$,
where $\lieder{}$ denotes the Lie derivative. 
This invariance holds for radial reduction. 
In $n=2$ dimensions, however, any vector field of the form $\vec{u}(r,t)$ is spherically symmetric. 

The equivalent equations of radial gas dynamics are given by 
the radial reduction of the pressure equation \eqref{pressure.eqn}  
\begin{equation}
p_t + U p_r + a^2 \rho (U_r +\tfrac{n-1}{r} U) = 0 , 
\label{radialgasdyn.p.eqn}
\end{equation}
and the velocity equation \eqref{velocity.eqn}
\begin{equation}
U_t + U U_r + p_r/\rho =0 .
\label{radialgasdyn.U.eqn}
\end{equation}
Here the sound speed \eqref{soundspeed} has the role of an equation of state,
through the correspondence \eqref{asq.rel}.

\subsection{Radial conservation laws}

A local conservation law of the fluid equations \eqref{velocity.eqn}--\eqref{entropy.eqn}
is a continuity equation 
\begin{equation}
\big( D_t \Phi^t + D_{\vec{x}}\cdot\vec{\Phi} \big)\big|_\Esp =0
\label{conslaw}
\end{equation}
holding on all solutions of the equations,
where $\Phi^t$ is the conserved density and $\vec{\Phi}$ is the spatial flux,
which are functions of the fluid variables and their spatial derivatives, in addition to $\vec{x}$ and $t$. 
In particular, note that all $t$-derivatives can be eliminated through use of equations \eqref{velocity.eqn}--\eqref{entropy.eqn}.
Here $\Esp$ denotes the space of solutions; 
$D_t$ denotes a total time derivative, 
and $D_{\vec{x}}$ denotes a total gradient (space derivative),
which are given by $\partial_t$ and $\nabla$ on functions of $\vec{x}$ and $t$,
and which act via the chain rule on the fluid variables and their spatial derivatives. 

Under radial reduction, 
the spatial flux is required to have the form $\vec\Phi = \Phi^r\rvec$, 
whereby a conservation law \eqref{conslaw} becomes 
\begin{equation}
\big( D_t \Phi^t + D_r \Phi^r + \tfrac{n-1}{r} \Phi^r \big)\big|_\Esp =0
\label{radial.reduc.conslaw}
\end{equation}
due to $\nabla\cdot \rvec = (n-1)/r$. 
A radial conservation law \eqref{radial.reduc.conslaw} can be expressed 
in an equivalent form of a total $t,r$-derivative
\begin{equation}
\big( D_t(r^{n-1} \Phi^t) + D_r(r^{n-1}\Phi^r) \big)\big|_\Esp =0 . 
\label{radial.conslaw}
\end{equation}
The conserved density $\Phi^t$ and radial flux $\Phi^r$ 
are functions of the fluid variables $U(r,t)$, $\rho(t,r)$, $S(t,r)$, 
and their radial derivatives, in addition to $r$ and $t$.

\subsection{Radial conserved integrals}

Every radial conservation law \eqref{radial.conslaw} can be integrated over $0\leq r<\infty$ 
to obtain a conserved (time-independent) radial integral
\begin{equation}
\frac{d}{dt}\int_0^\infty \Phi^t\,r^{n-1}dr\Big|_\Esp = 0
\end{equation}
if the radial flux satisfies the conditions 
$\lim_{r\to 0} r^{n-1} \Phi^r =0$ and $\lim_{r\to \infty} r^{n-1} \Phi^r =0$. 
The conserved integral $\int_0^\infty \Phi^t\,r^{n-1}dr\big|_\Esp$
is related to the radial reduction of an $n$-dimensional conserved integral 
which satisfies
$\frac{d}{dt}\int_{\Rnum^n} \Phi^t \,d^nx\big|_\Esp =-\lim_{r\to \infty}\oint_{S^n(r)} \vec{\Phi}\cdot\rvec\, d^{n-1}A$
from integration of a $n$-dimensional conservation law \eqref{conslaw} over $\Rnum^n$
followed by use of the divergence theorem, 
where $S^n(r)$ denotes the $n$-sphere of radius $r$
and $d^{n-1}A$ denotes the area element. 
Radial reduction leads to the relations
$\int_{\Rnum^n} \Phi^t \,d^nx =\vol(S^n) \int_0^\infty \Phi^t\,r^{n-1}dr$
and
$\oint_{S^n(r)} \vec{\Phi}\cdot\rvec\, d^{n-1}A = \vol(S^n) \Phi^r\,r^{n-1}$, 
where $S^n$ is the unit $n$-sphere and $\vol(S^n)$ denotes its hypersurface area. 

In general, a conservation law is locally trivial \cite{Olv-book,Anc-review,AncChe} 
when it holds as a differential identity that contains no local information about solutions.
Namely, the conserved density is given by a total spatial divergence 
$\Phi^t=D_{\vec{x}}\cdot \vec{\Theta}$, 
while the spatial flux is given by a time derivative 
$\vec{\Phi} = -D_t\vec{\Theta}|_\Esp$,
where $\vec{\Theta}$ is a vector function of the fluid variables and their spatial derivatives, in addition to $\vec{x}$ and $t$. 
Thus, in the case of a radial conservation law \eqref{radial.conslaw}, 
it is locally trivial if and only if 
\begin{equation}\label{loctriv}
\Phi^t =D_r\Theta +\tfrac{n-1}{r}\Theta,
\quad
\Phi^r =-D_t\Theta|_\Esp
\end{equation}
holds for some scalar function $\Theta$ of the radial fluid variables 
and their radial derivatives, in addition to $r$ and $t$. 
Only non-trivial radial conservation laws are of physical interest. 
Two radial conservation laws that differ by a locally trivial conservation law 
are physically equivalent. 

It is useful in fluid dynamics to consider conserved integrals on 
finite moving domains $V(t)$ that are transported by the flow in $\Rnum^n$
(see e.g.\ \cite{Ibr1973,Ibr-CRC,ArnKhe}). 
For radial flows, this type of conserved integral has the form 
\begin{equation}
\frac{d}{dt}\int_{V(t)} \Phi^t\,r^{n-1}dr\Big|_\Esp 
= -\left( r^{n-1} \Psi\right) \Big|_{\partial V(t)}
\label{moving.consintegral}
\end{equation}
whose physical content is that 
the rate of change of the moving integral $\int_{V(t)} \Phi^t\,r^{n-1}dr\big|_\Esp$ 
on a transported radial domain $V(t)$ 
is balanced by the net outward radial flux $(r^{n-1} \Psi)\big|_{\partial V(t)}$
through the moving boundary $\partial V(t)$. 
A moving conserved integral will hold for all transported radial domains
if and only if the conserved density $\Phi^t$ $(\equiv\mathrm{C})$ and moving radial flux $\Psi$ $(\equiv\vec{\mathrm{F}}\cdot\hat r)$
satisfy a local radial conservation law \eqref{radial.conslaw} in which 
$\Phi^r = \Psi + U\Phi^t$. 
Stated equivalently, any local radial conservation law \eqref{radial.conslaw} 
yields a moving conserved integral \eqref{moving.consintegral} 
on any transported radial domain $V(t)$, 
with the moving flux being given by 
\begin{equation}
\Psi = \Phi^r - U\Phi^t . 
\label{moving.flux}
\end{equation}

For all transported domains, 
a moving conserved integral \eqref{moving.consintegral} will yield a constant of motion,
$\frac{d}{dt}\int_{V(t)} \Phi^t\,r^{n-1}dr\big|_\Esp=0$, also called an integral invariant, 
if and only if $\Psi|_\Esp\equiv 0$. 

Note that the moving flux of a locally trivial radial conservation law is given by 
$\Psi|_\Esp = -(D_t \Theta + U D_r \Theta)|_\Esp = -\frac{d}{dt}\Theta|_\Esp$
in terms of the material derivative \eqref{materialder}. 
Hence, the corresponding moving conserved integral is simply an identity
$\frac{d}{dt}\int_{V(t)} D_r(r^{n-1}\Theta)\,dr\big|_\Esp 
= -( -r^{n-1}\frac{d}{dt}\Theta ) \big|_{\partial V(t)}$. 
Two moving conserved integrals that differ by a locally trivial moving conserved integral
are thereby physically equivalent.

\section{Invariants}\label{sec:invs}

An invariant of the $n$-dimensional Euler equations \eqref{velocity.eqn}--\eqref{entropy.eqn}
is a quantity $\J$ that is advected (frozen-in) by the flow. 
This property has the geometrical formulation that $\J$ is Lie dragged by the flow, 
\begin{equation}
\left( D_t\J + \lieder{\vec{u}}\J \right)\mid_\Esp =0
\end{equation}
where $\lieder{\vec{u}}$ is the Lie derivative with respect to the velocity $\vec{u}$.
The underlying operator $D_t + \lieder{\vec{u}}$ here
is a tensorial generalization of the material derivative 
$d/dt = \partial_t + \vec{u}\cdot\nabla$. 
In general, $\J$ is a function of the fluid variables, their spatial derivatives, and $\vec{x}$, $t$. 
The simplest example is entropy $\J=S$, which is a scalar invariant. 

For radial flow, 
as governed by the radial Euler equations \eqref{radialU.eqn}--\eqref{radialS.eqn},
invariants can be obtained by radial reduction of invariants in $\Rnum^n$. 
However, apart from $S$, all known ones \cite{ArnKhe,BesFri,AncWeb2020} involve vorticity 
and hence they become trivial in radial flows. 
Nevertheless, invariants may exist which are not inherited through reduction. 
Such quantities are sometimes called ``hidden''. 

There are three basic geometrical types of invariants: scalars, vectors, 1-forms. 
In radial flow, 
these types turn out to be mutually related to each other and to conservation laws. 

\subsubsection*{Scalars}
A scalar function $\J=J$ will be an invariant of the radial Euler equations \eqref{radialU.eqn}--\eqref{radialS.eqn} 
if and only if it satisfies
\begin{equation}
\frac{dJ}{dt}\Big|_\Esp =\left(D_t J + U D_r J\right)\mid_\Esp =0
\label{scalarinv.deteqn}
\end{equation}
since the Lie derivative acts as $\lieder{\vec{u}}=\vec{u}\cdot D_{\vec{x}} = U D_r$. 
Thus, a scalar invariant is simply an advected (material) quantity. 

Every scalar invariant can be easily shown to correspond to a local conservation law \eqref{radial.conslaw}
given by $\Phi^t = r^{n-1}\rho J$ and $\Phi^r = r^{n-1}\rho U J$.
Since $\Psi\equiv 0$, 
the conservation law yields an integral invariant $\int_{V(t)} \rho J\,r^{n-1} dr|_\Esp$. 

\subsubsection*{1-forms}
A radial 1-form can be expressed as $\J = J\d{r}$ in terms of a scalar function $J$. 
To yield a 1-form invariant of the radial Euler equations \eqref{radialU.eqn}--\eqref{radialS.eqn}, 
$J$ must satisfy 
\begin{equation}
\left(D_t J + D_r(U J)\right)\mid_\Esp =0 . 
\label{1forminv.deteqn}
\end{equation}
This follows from Cartan's formula for the action of a Lie derivative on a 1-form:
$\lieder{\vec{u}}\J = \vec{u}\hook\d{\J} + \d{(\vec{u}\hook\J)}$. 
Its radial version is obtained via the relations 
$\d{\J} = J_r\d{r}\wedge\d{r}=0$
and 
$\d{(\vec{u}\hook\J)} = \d{(UJ)} = D_r(UJ)\d{r}$. 

Every radial 1-form invariant directly corresponds to a local conservation law \eqref{radial.conslaw} with $\Phi^t = r^{1-n}J$ and $\Phi^r = r^{1-n}U J$. 
This yields an integral invariant $\int_{V(t)} J\,dr|_\Esp$, 
since the conservation law has $\Psi\equiv 0$. 

\subsubsection*{Vectors}
A radial vector can be expressed as $\J = J\rvec$ in terms of a scalar function $J$. 
For $J\rvec$ to be a vector invariant of the radial Euler equations \eqref{radialU.eqn}--\eqref{radialS.eqn}, 
$J$ must satisfy 
\begin{equation}
(D_t J + UD_r J -JD_r U)|_\Esp =0 . 
\label{vectorinv.deteqn}
\end{equation}
This is a consequence of the commutator formula for the action of a Lie derivative on a vector: 
$\lieder{\vec{u}}\J = \vec{u}\cdot D_{\vec{x}} \J - \J\cdot D_{\vec{x}} \vec{u}$. 
The radial version follows from the relations 
$\vec{u}\cdot D_{\vec{x}} \J = (U D_r J)\rvec$
and 
$\J\cdot D_{\vec{x}} \vec{u} = (J D_r U)\rvec$
along with $D_r \rvec=0$.

\subsection{Properties}

These three basic types of invariants are mutually related through their
determining equations \eqref{scalarinv.deteqn}, \eqref{1forminv.deteqn}, \eqref{vectorinv.deteqn}
in terms of $J$. 

\begin{proposition}\label{prop:inv.relations}
The following conditions are equivalent:
\\
(i) $J$ is a scalar invariant;
\\
(ii) $r^{n-1}\rho J\d{r}$ is a 1-form invariant;
\\
(iii) $r^{1-n}/(\rho J)\rvec $ is a vector invariant;
\\
(iv) $\int_{V(t)} \rho J\,r^{n-1} dr|_\Esp$ is an integral invariant for all transported radial domains $V(t)$. 
\end{proposition}

The proof amounts to verifying that 
if $J_{\text{scal.}}=J$ satisfies equation \eqref{scalarinv.deteqn}
then $J_{\text{1-form}} = r^{n-1}\rho J$ satisfies equation \eqref{1forminv.deteqn}
and $J_{\text{vec.}} = r^{1-n}/(\rho J)$ satisfies equation \eqref{vectorinv.deteqn}, 
and conversely. 

A similar computation shows that 
$J_{\text{vec.}}D_r J_{\text{scal.}} =J$ 
satisfies equation \eqref{scalarinv.deteqn}. 
This yields the following interesting result which provides 
a recursion operator on invariants. 

\begin{proposition}\label{prop:inv.recursion}
Suppose $J_1$ and $J_2$ are scalar invariants. 
Then: (i) $J_3 = (r^{1-n}/\rho) J_2 D_r J_1$ and $J_4=f(J_1,J_2)$ are also scalar invariants,
where $f$ is any function on $\Rnum^2$;
(ii) $\I =\int_{V(t)} J_2(D_r J_1)\,dr|_\Esp$ is an integral invariant,
where $V(t)$ is any transported radial domain. 
This integral is non-trivial iff $D_r(J_2/J_1)\not\equiv 0$. 
\end{proposition}

Corresponding versions can be formulated for 1-form invariants and vector invariants.

\subsection{Results}

All invariants can be found, in principle, by solving one of the determining equations
\eqref{scalarinv.deteqn}, \eqref{1forminv.deteqn}, \eqref{vectorinv.deteqn}. 
A classification of scalar invariants up to first-order, 
$J(t,r,U,\rho,S,U_r,\rho_r,S_r)$, 
will now be presented.
Remarks on the computation are given in Appendix~\ref{app:invs.computation}.

\begin{theorem}\label{thm:1storder.invs}
(i) For a general \eos/ \eqref{eos}, 
all scalar invariants up to first-order are functions of 
\begin{equation}
S,
\quad
r^{1-n}S_r/\rho . 
\end{equation}
(ii) The only special \eos/ for which additional scalar invariants arise up to first-order 
is the entropic case $p=\k(S)$,
where $\k$ is an arbitrary non-constant function. 
The additional invariants are functions of 
\begin{equation}
U^2 +\tfrac{2}{n}r \k'(S) S_r/\rho, 
\quad
\int_0^r \frac{dy}{\sqrt{U^2 +\tfrac{2}{n}(1-(y/r)^n) r \k'(S) S_r/\rho}} - t . 
\label{entropic.1storder.invs}
\end{equation}
\end{theorem}

From part (i) of Proposition~\ref{prop:inv.recursion}, 
the trivial invariant $J=1$ yields a recursion operator 
\begin{equation}
\Rop = (r^{1-n}/\rho) D_r . 
\label{inv.recursionop}
\end{equation}
When this operator is applied to the basic invariant $S$, 
the first-order invariant $r^{1-n}S_r/\rho$ is obtained. 
Repeated application yields a hierarchy of successively higher-order invariants,
\begin{equation}
J_l = \Rop^l S
\label{J.hierarchy}
\end{equation}
for $l=1,2,\ldots$, 
with the second-order invariant being given by 
$(r^{1-n}/\rho)^2( S_{rr} -S_r\rho_r/\rho + \frac{1-n}{r} S_r )$.
Moreover,  
all second-order invariants for a general \eos/ \eqref{eos} 
turn out to be exhausted by an arbitrary function of $S$, $J_1$, $J_2$,
as discussed in Appendix~\ref{app:invs.computation}. 

In the case of an entropic \eos/ $p=\k(S)$,
additional higher-order invariants can be obtained by applying $\Rop$ 
to the two invariants \eqref{entropic.1storder.invs}:
\begin{align}
& J_{1,l}=\Rop^{l-1}\big( U^2 +\tfrac{2}{n}r p_r/\rho \big) , 
\label{J1.hierarchy}
\\
& 
J_{2,l} = \Rop^{l-1}\big( A(r,U,p_r/\rho) - t \big)
= \int_0^1 \Rop^l\Big(r/\sqrt{U^2 +\tfrac{2}{n}(1-y^n) r p_r/\rho}\Big) dy , 
\label{J2.hierarchy}
\end{align}
for $l=2,3,\ldots$,
where
\begin{equation}
A(r,U,p_r/\rho) = \int_0^r \frac{dy}{\sqrt{U^2 + \tfrac{2}{n}(1-(y/r)^n)r p_r/\rho}}
= \int_0^1 \frac{r\,dy}{\sqrt{U^2 +\tfrac{2}{n}(1-y^n) r p_r/\rho}}
\label{A}
\end{equation}

Thus, the following main result has been established. 

\begin{theorem}\label{thm:invs.hierarchies}
The radial Euler equations \eqref{radialU.eqn}--\eqref{radialS.eqn}
possess a hierarchy of scalar invariants \eqref{J.hierarchy} of order $l=0,1,2,\ldots$
for a general \eos/, 
and two additional hierarchies of scalar invariants \eqref{entropic.1storder.invs} of order $1$, \eqref{J1.hierarchy} and \eqref{J2.hierarchy} of order $l=2,3,\ldots$, 
for an entropic \eos/. 
\end{theorem}

Finally, 
each hierarchy of invariants yields a corresponding hierarchy of invariant integrals, 
which will be discussed in the next section.

\section{Conserved integrals}\label{sec:conslaws}

It will be useful to divide the set of non-trivial conserved integrals into two distinct types:
ones whose moving flux is zero, which are integral invariants representing advected quantities; 
ones with non-zero moving flux, which represent non-advected conserved quantities. 
For each type, a classification result will be presented. 

These classifications will show that 
the radial Euler equations \eqref{radialU.eqn}--\eqref{radialS.eqn} 
possess both types of conserved integrals besides those that are inherited from
radial reduction of the conserved integrals known for $n$-dimensional fluid flow
\cite{Ibr1973,Ibr-CRC,AncDar2009,AncDar2010}. 
The known conserved integrals comprise kinematic conservation laws ---
whose conserved density and spatial flux involve only 
the time and space coordinates, and the fluid variables, but not their derivatives;
and vorticity conservation laws --- 
in which the conserved density involves the vorticity scalar in even dimensions
and the vorticity vector in odd dimensions \cite{ArnKhe}.
A complete classification is shown in \Ref{AncDar2010},
which encompasses all \esos/ \eqref{eos} excluding the entropic case $p=\k(S)$. 
Under radial reduction, the vorticity conserved integrals are trivial, 
since $\nabla\wedge\vec{u}=0$.

\subsection{Integral invariants}\label{sec:invintegrals}

The classification of scalar invariants (up to first order) stated in Theorem~\ref{thm:1storder.invs}
provides, through Proposition~\ref{prop:inv.relations}, 
a corresponding classification of radial integral invariants 
\begin{equation}
\frac{d}{dt} \int_{V(t)} \Phi^t\,r^{n-1} dr\Big|_\Esp =0
\label{integral.inv}
\end{equation}
with conserved densities of the form 
\begin{equation}
\Phi^t = \rho J(t,r,U,\rho,S,U_r,\rho_r,S_r) . 
\label{1storder.consdens}
\end{equation}
These conserved integrals are advected quantities for 
the radial Euler equations \eqref{radialU.eqn}--\eqref{radialS.eqn}. 

\begin{theorem}\label{thm:1storderintegral.invs}
(i) For a general \eos/ \eqref{eos}, 
all integral invariants \eqref{integral.inv} up to first order \eqref{1storder.consdens} 
are given by 
\begin{equation}
\frac{d}{dt} \int_{V(t)} \rho f(S,r^{1-n}S_r/\rho) \,r^{n-1} dr\Big|_\Esp =0
\label{1storder.integral.invs}
\end{equation}
where $f$ is an arbitrary function. 
(ii) The only special \eos/ for which additional integral invariants arise up to first order 
is the entropic case $p=\k(S)$,
where $\k$ is an arbitrary non-constant function. 
The additional integral invariants are given by 
\begin{equation}
\frac{d}{dt} \int_{V(t)} \rho f\bigg(
S, r^{1-n}S_r/\rho, U^2 +\tfrac{2}{n}r p_r/\rho, A(r,U,p_r/\rho) -t
\bigg)\,r^{n-1} dr\Big|_\Esp =0
\label{entropic.1storderintegral.invs}
\end{equation}
where $f$ is an arbitrary function,
and $A(r,U,p_r/\rho)$ is expression \eqref{A}.  
\end{theorem}

The integral invariants of zeroth order comprise 
mass $\int_{V(t)} \rho\,r^{n-1} dr$ when $f=1$, 
and total entropy in the generalized form 
$\int_{V(t)} \rho f(S)\,r^{n-1} dr$ when $f$ is non-constant. 
These two advected quantities are inherited 
from radial reduction of the kinematic integral invariants 
known for the $n$-dimensional Euler equations \eqref{velocity.eqn}--\eqref{entropy.eqn}
from the classification of kinematic conservation laws in \Ref{AncDar2010}. 

The first-order integral invariants in both cases (i) and (ii) are new. 
They can be viewed as ``hidden'' advected quantities which exist only for radial flow. 

In case (i), 
the integral \eqref{1storder.integral.invs} is non-trivial 
whenever $f$ is not a linear homogeneous function in its second argument,
namely $f(S,J_1) \neq F(S) J_1$, 
since otherwise the conserved density has the locally trivial form 
$r^{n-1} \Phi^t = F(S) S_r$ which is a total radial derivative. 
As an example, the entropy-gradient integrals 
\begin{equation}\label{gradS.q.integral}
\int_{V(t)} \rho \big(r^{1-n}S_r/\rho\big)^{2q}\,r^{n-1} dr
= \int_{V(t)} \rho^{1-2q} |S_r|^{2q}\,r^{(n-1)(1-2q)} dr
\quad
q=1,2,\ldots
\end{equation}
are non-negative advected quantities for any \eos/.

In case (ii), 
the integral \eqref{entropic.1storderintegral.invs} is non-trivial 
whenever $f$ is a non-constant function in at least one of its last two arguments, 
namely it has some dependence on $U$. 
As an example, the energy-like integrals 
\begin{equation}\label{ener.q.integral}
\int_{V(t)} \rho \big(\tfrac{1}{2} U^2 +\tfrac{r}{n} p_r\big)^q\,r^{n-1} dr,
\quad
q=1,2,\ldots
\end{equation}
are advected for an entropic \eos/. 
For $i=1$, note that this integral is equivalent to the energy integral \eqref{energy.integral} 
with $p=\k(S) = -\rho e$. 
The equivalence can be seen from the relation 
$\tfrac{1}{2} U^2 +\tfrac{1}{n}r p_r/\rho = r^{n-1}( \tfrac{1}{2}\rho U^2 \rho e ) + D_r (\tfrac{1}{n}r^n p)$ 
between the densities, 
which differ by a total $r$-derivative, 
so thus the two integrals agree modulo a trivial integral. 

Another example is the integral quantity 
\begin{equation}\label{A.integral}
\int_{V(t)} \rho (A(r,U,p_r/\rho) - t)\,r^{n-1} dr = \A_{V(t)} - t  \M_{V(t)},
\end{equation}
which has explicit dependence on $t$,
where $\M_{V(t)}=\int_{V(t)} \rho\,r^{n-1} dr$ is the mass integral,
and $\A_{V(t)} = \int_{V(t)} \rho A(r,U,p_r/\rho)\,r^{n-1} dr$ is a non-conserved integral. 
Note, as a consequence, 
\begin{equation}\label{A.integral.increasing}
\frac{d}{dt} \A_{V(t)} = M_{V(t)}=\text{const.} \geq 0
\end{equation}
implies that $\A_{V(t)}$ is a non-decreasing quantity in the flow.

In both cases (i) and (ii), 
integral invariants of higher-order can be obtained from 
the hierarchies of scalar invariants \eqref{J.hierarchy}, \eqref{J1.hierarchy} and \eqref{J2.hierarchy}. 

\begin{theorem}\label{thm:integral.invs.hierarchy}
For any \eos/, 
\begin{equation}\label{integral.inv.Jhierarchy}
\I_l = \int_{V(t)} \rho f(J_0,J_1,\ldots,J_l) \,r^{n-1} dr
\end{equation}
is an integral invariant of order $l\geq 1$ if $f$ is non-constant in its last argument. 
It is non-trivial at order $l$ if and only if $f$ is nonlinear in its last argument, 
namely $f_{J_l J_l}\not\equiv0$.
\end{theorem}

The demonstration of non-triviality follows from expressing the conserved density as
\begin{equation}
r^{n-1} \Phi^t= F D_r J_{l-1} 
= D_r\big( \smallint F\,d J_{l-1} \big)  
- r^{n-1}\rho \smallint( J_1 F_{J_0} + \cdots + J_{l-1} F_{J_{l-2}} )\,d J_{l-1}
\end{equation}
which is locally trivial modulo terms of order less than $l$. 

A similar result holds for an entropic \eos/,
where $f$ has additional dependence on the higher-order invariants \eqref{J1.hierarchy}, \eqref{J2.hierarchy}, 
and as well as the two first-order invariants \eqref{entropic.1storder.invs}, 
denoted $J_{1,1}$ and $J_{2,1}$ respectively. 

\begin{theorem}\label{thm:integral.invs.hierarchy.entropic}
For an entropic \eos/,
\begin{equation}\label{integral.inv.J1J2hierarchy}
\I'_l = \int_{V(t)} \rho f(J_0,J_1,J_{1,1},J_{2,1},\ldots,J_l,J_{1,l},J_{2,l}) \,r^{n-1} dr
\end{equation}
is an integral invariant of order $l\geq 1$ if $f$ is non-constant in at least one of its last three arguments. 
It is non-trivial at order $l\geq2$ if and only if $f$ is nonlinear in at least one of $J_{1,l}$ and $J_{2,l}$; 
at order $l=1$, it is non-trivial if and only if $f$ is non-constant in at least one of $J_{1,1}$ and $J_{2,1}$. 
\end{theorem}

The question of whether these two hierarchies of integral invariants
exhaust all possible integral invariants of higher-order is much harder problem
which will be considered elsewhere.

\subsection{Non-advected conserved integrals}\label{sec:noninvintegrals}

Consider a conserved integral with non-zero moving flux: 
\begin{equation}
\frac{d}{dt}\int_{V(t)} \Phi^t \,r^{n-1} dr 
= - ( r^{n-1} \Psi )\Big|_{\partial V(t)},
\quad
\Psi = \Phi^r - U\Phi^t \not\equiv 0 . 
\label{consintegral}
\end{equation}
Here the conserved density $\Phi^t$ and radial flux $\Phi^r$ are separate functions,
in contrast to the form $\Phi^r = U\Phi^t$ characterizing integral invariants.

The first aim will be to find all conserved integrals given by 
kinematic radial conservation laws 
\begin{equation}
\Phi^t(t,r,U,\rho,S), 
\quad
\Phi^r(t,r,U,\rho,S) = \Psi(t,r,U,\rho,S) + U\Phi^t
\label{kinematic}
\end{equation}
in which the moving flux $\Psi$ is non-zero. 

The classification of kinematic conservation laws in \Ref{AncDar2010}
for the $n$-dimensional Euler equations \eqref{velocity.eqn}--\eqref{entropy.eqn}
shows that all kinematic conserved integrals having non-zero moving flux 
are a linear combination given by 
momentum, angular momentum, Galilean momentum, and energy 
for a general \eos/, 
and also by a dilational energy and a similarity energy 
for a polytropic \eos/. 
Their radial reduction yields the following non-trivial conserved integrals
for the radial Euler equations \eqref{radialU.eqn}--\eqref{radialS.eqn}:
\begin{align}
& \text{energy}\quad
\frac{d}{dt}\int_{V(t)} \rho (\tfrac{1}{2} U^2 +e) \,r^{n-1} dr 
= - ( r^{n-1} p U )\Big|_{\partial V(t)}
\label{energy.integral}
\end{align}
in the case of a general \eos/ $p=P(\rho,S)$, 
where $e=\int P(\rho,S)/\rho^2\, d\rho$;
and 
\begin{align}
& \text{dilational energy}\quad
\frac{d}{dt} \int_{V(t)} \big( t \rho (\tfrac{1}{2} U^2 +e) -\tfrac{1}{2} r \rho U \big) \,r^{n-1} dr
= - \big( r^{n-1} ( t  U -\tfrac{1}{2} r ) p \big)\Big|_{\partial V(t)} , 
\label{dilenergy.integral}
\\
& \text{similarity energy}\quad
\frac{d}{dt} \int_{V(t)} \big( t^2 \rho (\tfrac{1}{2} U^2 +e) -t\, r \rho U  +\tfrac{1}{2} r^2 \rho \big)\,r^{n-1} dr
= - \big( r^{n-1} t ( t  U - r ) p \big)\Big|_{\partial V(t)}
\label{simenergy.integral}
\end{align}  
in the case of an ideal polytropic \eos/ $p=\k(S)\rho^{1+2/n}$,
where $e=\tfrac{n}{2}\k(S) \rho^{2/n}$
and $\k$ is an arbitrary function. 
The radial reduction of angular momentum vanishes 
while momentum and Galilean momentum have no radial reduction. 

Additional (new) kinematic conserved integrals \eqref{consintegral}
can be sought by formulating and solving determining equations 
for conserved densities. 

Any radial conservation law \eqref{radial.conslaw} can be expressed as 
\begin{equation}\label{Phi.eqn}
D_t(r^{n-1} \Phi^t)|_\Esp = -D_r(r^{n-1}\Phi^r)
\end{equation}
where the conserved density $\Phi^t$ and the radial flux $\Phi^r$ 
contain no $t$-derivatives of the fluid variables. 
Since the righthand side of this equation is a total radial derivative, 
a function $\Phi^t$ with no $t$-derivatives of the fluid variables
will be a conserved density if and only if it satisfies the variational conditions
\begin{equation}
\delta (r^{n-1} D_t \Phi^t)|_\Esp/\delta U =0,
\quad
\delta (r^{n-1} D_t \Phi^t)|_\Esp/\delta \rho =0,
\quad
\delta (r^{n-1} D_t \Phi^t)|_\Esp/\delta S =0. 
\label{deteqns}
\end{equation}
These conditions constitute a set of linear determining equations 
for finding conserved densities modulo a locally trivial density 
$r^{n-1}\Phi^t = D_r \Theta$. 
This freedom in solutions disappears for kinematic conservation laws \eqref{kinematic}.
Once a solution for $\Phi^t$ has been found, 
then $\Phi^r$ can be determined from equation \eqref{Phi.eqn}
by inverting the total $r$-derivative. 

The determining equations \eqref{deteqns} for kinematic conserved densities 
each split with respect to $r$-derivatives of $U$, $\rho$, $S$,
thereby yielding an overdetermined system of PDEs for 
$\Phi^t$, $p(\rho,S)$, and $n$ as unknowns. 
This system is computationally straightforward to solve
and yields the following classification result. 
Remarks on the computation are given in Appendix~\ref{app:consdens.computation}. 

\begin{theorem}\label{thm:kinematic.consintegrals}
(i) For a general \eos/ \eqref{eos}, 
energy \eqref{energy.integral} is the only 
kinematic conserved integral \eqref {consintegral}--\eqref{kinematic} 
with non-zero moving flux $\Psi\not\equiv 0$. 
(ii) The only special \esos/ for which additional non-advected kinematic conserved integrals exist 
are the ideal polytropic case $p=\k(S)\rho^{1+2/n}$,
the barotropic case $p=P(\rho)$, 
and the entropic case $p=\k(S)$. 
The additional conserved integrals are respectively given by:
dilational energy \eqref{dilenergy.integral} and similarity energy \eqref{simenergy.integral}
for an ideal polytropic \eos/;
\begin{equation}
\text{enthalpy flux}\quad
\frac{d}{dt} \int_{V(t)} U\, dr\Big|_\Esp 
= - ( e + p/\rho - \tfrac{1}{2} U^2 )\Big|_{\partial V(t)}
\label{enthaplyflux.integral}
\end{equation}
for a barotropic \eos/,
where $e=\int p(\rho)/\rho^2\, d\rho$; 
\begin{equation}
\text{entropy-weighted energy}\quad
\frac{d}{dt} \int_{V(t)} \big( \tfrac{1}{2}\rho U^2 F(S) - K(S) \big)\, r^{n-1} dr\Big|_\Esp 
= -\big( r^{n-1} U K(S) \big)\Big|_{\partial V(t)}
\label{nonisentropic.energy.integral}
\end{equation}
for an entropic \eos/,
where $e=-\k(S)/\rho$ and $K(S)=\int F(S) \k'(S)\,dS$, 
with $F$ being an arbitrary non-constant function. 
\end{theorem}

The entropy-weighted energy \eqref{nonisentropic.energy.integral} is 
the radial reduction of an analogous conserved integral that was first found 
in a classification of kinematic conserved integrals for the Euler equations formulated
on $n$-dimensional Riemannian manifolds \cite{AncDarTuf}.  
It can be written in the equivalent form 
\begin{equation}
\frac{d}{dt} \int_{V(t)} \big( (\tfrac{1}{2} U^2 +e)\rho F(S) +\smallint F'(S)p\,dS \big)\, r^{n-1} dr\Big|_\Esp 
= -\big( \tfrac{1}{2} r^{n-1} U ( pF(S) -\smallint F'(S)p\,dS ) \big)\Big|_{\partial V(t)}
\label{nonisentropic.energy.integral.alt}
\end{equation}
which shows how it reduces to the energy integral \eqref{energy.integral} 
when $f(S)=1$. 

The possibility that new higher-order conserved integrals with non-zero moving flux
may exist is suggested by the existence of the new first-order scalar invariants. 
To-date, no classification of higher-order conserved integrals 
for the $n$-dimensional Euler equations \eqref{velocity.eqn}--\eqref{entropy.eqn} 
has been carried out, beyond the results in \Ref{AncDar2009,AncDar2010,AncDarTuf}
for vorticity conservation laws. 

Here, a systematic search for such conserved integrals \eqref{consintegral} 
for the radial Euler equations \eqref{radialU.eqn}--\eqref{radialS.eqn}
will be undertaken by solving the determining equations \eqref{deteqns}
for first-order conserved densities. 
This general problem turns out to be a very difficult computation. 
Consequently, attention will be restricted to first-order conserved densities that are
quadratic in $U_r$, with no explicit dependence on $t$:
\begin{equation}
\Phi^t= U_r{}^2 \phi_2(r,U,\rho,S,\rho_r,S_r) + U_r\phi_1(r,U,\rho,S,\rho_r,S_r) + \phi_0(r,U,\rho,S,\rho_r,S_r)
\label{1storder.consdens.quadr.Ur}
\end{equation}
where at least one of $\phi_2$ and $\phi_1$ is not identically zero. 

The determining equations 
each split with respect to $r$-derivatives of $U_r$, $\rho_r$, $S_r$,
similarly to the case for kinematic conserved densities. 
By use of various lengthy steps, combining several integration techniques, 
the resulting large overdetermined system of PDEs 
for the unknowns $\Phi^t$, $p(\rho,S)$, and $n$ can be solved, 
modulo locally trivial conserved densities. 
Computational remarks are given in Appendix~\ref{app:consdens.computation}

This yields the following classification result. 

\begin{theorem}\label{thm:1storder.consintegrals}
(i) For a general \eos/ \eqref{eos}, 
all first-order conserved densities of the form \eqref{1storder.consdens.quadr.Ur} 
are given by 
$\Phi^t = \rho f(S,J_1)$ 
modulo trivial densities,
where $f$ is an arbitrary function on $\Rnum^2$. 
(ii) The only special \esos/ for which there exist 
non-trivial first-order conserved densities of the form \eqref{1storder.consdens.quadr.Ur} 
is the entropic case $p=\k(S)$. 
The conserved densities are, modulo trivial densities,  
a linear combination given by:
\begin{equation}\label{newdens1}
\Phi^t_1 = 
r^{1-n}\int_0^{J'_1} D_r f(S,y,U^2 +\tfrac{2}{n}r^n y)\,dy
\end{equation}
and 
\begin{equation}\label{newdens2}
\begin{aligned}
\Phi^t_2 = 
r^{1-n}\Big(
f(S,J'_1,J_{1,1})/U
+ & \int_0^{J'_1} \int_0^r 
\big(  
(D_r (U^2 +\tfrac{2}{n}(r^n-z^n)y)^{-1/2})\partial_y f(S,y,U^2 +\tfrac{2}{n}r^n y) 
\\&
- (\partial_y (U^2 +\tfrac{2}{n}(r^n-z^n)y)^{-1/2}) D_r f(S,y,U^2 +\tfrac{2}{n}r^n y) 
\big)\,dz\,dy \Big)
\end{aligned}
\end{equation}
where $J'_1 = r^{1-n} p_r/\rho = \k'(S) J_1$ is a scalar invariant, 
and $f$ is an arbitrary differentiable function on $\Rnum^3$. 
\end{theorem}

Both of these non-trivial conserved densities turn out to be locally equivalent to 
conserved densities with zero moving flux: 
\begin{equation}
\Phi^t_1 = 
-\rho f(S,J'_1,J_{1,1}) +D_r\Theta_1
\end{equation}
and 
\begin{equation}
\Phi^t_2 = 
r^{1-n}f(S,J'_1,J_{1,1})D_r A(r,U,p_r/\rho) +D_r\Theta_2
\end{equation}
where $A(r,U,p_r/\rho)$ is expression \eqref{A}. 
This leads to the following classification result. 

\begin{corollary}\label{cor:1storder.consintegrals}
For any \eos/, 
every conserved integral \eqref {consintegral} 
given by a non-trivial first-order conserved density of the form \eqref{1storder.consdens.quadr.Ur} 
(modulo a trivial conserved density) is equivalent to an integral invariant. 
\end{corollary}

As a consequence, unlike the situation for integral invariants, 
no first-order recursion operator apparently exists in general for conserved densities
whose moving flux is non-zero.

\subsection{Scaling properties and applications}\label{sec:scaling}

It is worthwhile to examine the scaling behaviour of the conserved integrals 
\eqref{gradS.q.integral}, \eqref{ener.q.integral}, \eqref{A.integral}, 
\eqref{energy.integral}, \eqref{dilenergy.integral}, \eqref{simenergy.integral}
\eqref{enthaplyflux.integral}, \eqref{nonisentropic.energy.integral}. 
One application is that scaling of conserved energies is useful to know 
for study of the Cauchy problem in Sobolev spaces, 
or in energy spaces where an energy integral provides a conserved norm. 
Another application is to seek conserved integrals that will be inherited 
under scaling reduction when self-similar flows are considered. 

As shown by the results in \Ref{AncSeiWol}, 
the radial Euler equations \eqref{radialU.eqn}--\eqref{radialS.eqn}
possess several scaling symmetries for different \esos/ with $p\neq$const., 
which are listed in Table~\ref{table:scalingsymms}. 

\begin{table}[h]
\hbox{\hspace{-0.3in}
\begin{tabular}{l|c||c|c}
\hline
\qquad $p$
& scaling transformation
& parameters
& type 
\\
\hline
\hline
$P(\rho,S)$
& $t\to\lambda t$, $r\to\lambda r$
& 
& space-time 
\\
&&
& dilation 
\\
\hline
$S^\nu P(\rho)$
& $t\to\lambda^\alpha t$, $r\to\lambda^{\alpha+\nu/2} r$, $U\to\lambda^{\nu/2} U$, $S\to\lambda S$
& $\alpha$, $\nu$
& entropy 
\\
$\nu=$const.
&& 
& scaling
\\
\hline
$\k(S/\rho^\nu)\rho^{1+\gamma}$
& $t\to\lambda^\alpha t$, $r\to\lambda^{\alpha+\gamma/2} r$, $U\to\lambda^{\gamma/2} U$, $\rho\to\lambda \rho$, $S\to \lambda^{\nu} S$
& $\alpha$, $\gamma$, $\nu$
& density 
\\
$\gamma,\nu=$const.
&&
& scaling
\\
\hline
$\k_0 S^\nu$
& $t\to\lambda^\alpha t$, $r\to\lambda^{\alpha+\beta/2} r$, $U\to\lambda^{\beta/2} U$, $\rho\to\lambda^{\nu-\beta} \rho$, $S\to \lambda S$
& $\alpha$, $\beta$, $\nu$
& similarity
\\
$\k_0,\nu=$const.
&&
& 
\\
\hline
\hline
\end{tabular}
}
\caption{Scaling symmetries: 
$\lambda=\text{const.}\neq 0$.}
\label{table:scalingsymms}
\end{table}

If a conserved integral transforms homogeneously as $\lambda^w$ 
under a scaling transformation, 
then the power $w$ is its corresponding scaling weight. 
For each type of scaling in Table~\ref{table:scalingsymms}, 
the scaling weights of the kinematic conserved integrals 
and the first-order integral invariants 
are summarized respectively in 
Tables~\ref{table:scaling.kinematic} and~~\ref{table:scaling.invs}.
Note: ``---'' means that the scaling is incompatible with the \eos/ 
for which the conserved integral exists, 
allowing at most a specialization of the exponent(s) in the \eos/.

\begin{table}[h]
\hbox{\hspace{-0.5in}
\begin{tabular}{l||c|c|c|c}
\hline
conserved integral
& space-time 
& entropy
& density
& similarity
\\
& dilation
& scaling
& scaling
&
\\
\hline
\hline
energy \eqref{energy.integral}
& $n$
& $n\alpha + (1+\tfrac{1}{2}n)\nu$
& $n\alpha + (1+\tfrac{1}{2}n)\gamma +1$
& $n\alpha + \tfrac{1}{2}n\beta +\nu$
\\
\hline
dilational energy \eqref{dilenergy.integral}
& $n+1$
& ---
& $(n+1)(\alpha +2/n)$
& ---
\\
similarity energy \eqref{simenergy.integral}
& $n+2$
& ---
& $(n+2)\alpha +2(1+1/n)$
& ---
\\
& 
& 
& ($\gamma=2/n, \nu=0$)
& 
\\
\hline
enthalpy flux \eqref{enthaplyflux.integral}
& $1$
& $\alpha$
& $\alpha+\gamma$
& ---
\\
&  
& ($\nu=0$)
& ($\kappa=$const.)
& 
\\
\hline
entropy-weighted 
& $n$
& ---
& $n(\alpha -\tfrac{1}{2}) + \mu$
& $n\alpha +\tfrac{1}{2}n\beta + \mu +\nu$
\\
energy \eqref{nonisentropic.energy.integral}
&&
& ($\nu=0,\gamma=-1$)
&
\\
with $F(S)=F_0 S^\mu$
&&&&
\\
\hline
\hline
\end{tabular}
}
\caption{Scaling weights $w$ of kinematic conserved integrals.}
\label{table:scaling.kinematic}
\end{table}


\begin{table}[h]
\hbox{\hspace{-0.5in}
\begin{tabular}{l||c|c|c|c}
\hline
integral
& space-time 
& entropy 
& density
& similarity
\\
invariant 
& dilation
& scaling
& scaling
&
\\
\hline
\hline
entropy-gradient
& $(1-2q)n$
& $(1-2q)n(\alpha +\tfrac{1}{2}\nu) +2q$
& $(1-2q)n(\alpha +\tfrac{1}{2}\gamma)$
& $(1-2q)(n\alpha +(\tfrac{1}{2}n-1)\beta)$
\\
\eqref{gradS.q.integral}, $q=1,2,\ldots$
&
& 
& $\quad +2q(\nu-1)+1$
& $\quad +2q(1-\nu)+\nu$
\\
\hline
energy-like
& $n$
& ---
& $n(\alpha -\tfrac{1}{2}) + 1-2q$
& $n\alpha +(\tfrac{1}{2}n +2q-1)\beta +\nu$
\\
\eqref{ener.q.integral}, $q=1,2,\ldots$
&& 
& ($\nu=0,\gamma=-1$)
&
\\
\hline
non-decreasing
& $n+1$
& ---
& $(n+1)\alpha +1-\tfrac{1}{2}n$
& $(n+1)\alpha +(\tfrac{1}{2}n-1)\beta +\nu$
\\
\eqref{A.integral}
&&
& ($\nu=0,\gamma=-1$)
&
\\
\hline
\hline
\end{tabular}
}
\caption{Scaling weights $w$ of first-order integral invariants.}
\label{table:scaling.invs}
\end{table}

Two main observations from the tables are that 
each conserved integral is invariant under some scaling other than space-time dilation,
while the entropy-gradient integrals \eqref{gradS.q.integral} are subcritical 
and all of the other conserved integrals are supercritical under space-time dilation. 
There are several implications. 

Firstly, in regard to the Cauchy problem, 
compressible flows are well known to exhibit shocks and blow ups. 
If smooth solutions that have a finite lifespan are considered,
then the first observation shows that there are scaled solutions that have 
a longer or shorter lifespan but whose energy and other conserved quantities 
remain the same. 
Hence, the lifespan cannot be proportional to any non-zero power of these quantities. 
The second observation indicates that the entropy-gradient integrals \eqref{gradS.q.integral} 
may provide useful a priori estimates for non-barotropic flows in which entropy plays 
a significant role in the dynamics. 
In particular, they yield a weighted $L^{2q}$ norm for $S_r$, which is conserved 
for every $q=1,2,\ldots$. 

Secondly, for the study of shock formation, 
the quantity \eqref{A} associated to the conserved integral \eqref{A.integral} 
may be very useful. 
Because it is non-decreasing \eqref{A.integral.increasing}, 
it represents a new ``entropy'' \cite{NovStr} which is connected to the conserved mass 
in a compressible flow 
and differs from the physical entropy $S$. 

Finally, each of the four scaling symmetries listed in Table~\ref{table:scalingsymms}
can be used to define a corresponding self-similar flow
whose variables are given by the invariants of the scaling. 
For these flows, another implication of the first observation is that 
the local conservation laws \eqref{radial.conslaw} underlying the conserved integrals 
will be inherited as ODEs when a scaling leaves invariant the conserved integral. 
Furthermore, under certain conditions, these ODEs will take the form of first integrals 
which provide conserved quantities for the self-similar flows. 
The criteria for a conserved integral to reduce to such a conserved quantity 
is given by the theory of symmetry multi-reduction \cite{AncGan2020}
applied to the scaling symmetries that characterize self-similar flows. 
It will be shown in subsequent work \cite{Anc2023} that the reduction criteria hold
when the parameter $\alpha$ in the entropy scaling, density scaling, and similarity scaling 
satisfies a certain algebraic condition involving the dimension $n$ and the exponent(s)
in the corresponding \eos/. 
In particular, the new first-order conserved integrals 
\eqref{gradS.q.integral}, \eqref{ener.q.integral}, \eqref{A.integral} 
and the new kinematic conserved integrals 
\eqref{enthaplyflux.integral}, \eqref{nonisentropic.energy.integral},
along with the well-known conserved integrals 
\eqref{energy.integral}, \eqref{dilenergy.integral}, \eqref{simenergy.integral}, 
yield first integrals for the resulting self-similar flows.

\section{Concluding remarks}\label{sec:remarks}

Radial fluid flow possesses two additional kinematic conserved integrals 
apart from the well-known ones inherited under radial reduction of $n$-dimensional (non-radial) fluid flow. 
The inherited kinematic conserved integrals consist of 
total entropy, mass, and energy holding for a general \eos/,
as well as dilational energy and similarity energy holding for ideal polytropic \esos/. 
The new radial conserved integrals represent 
an enthalpy-flux quantity \eqref{enthaplyflux.integral} 
which holds for barotropic \esos/,
and an entropy-weighted energy \eqref{nonisentropic.energy.integral}
which holds for entropic \esos/. 

Most interestingly, 
radial fluid flow with a general \eos/ also possesses 
a hierarchy of integral invariants \eqref{integral.inv.Jhierarchy}
which describe advected conserved integrals. 
These quantities arise from a corresponding hierarchy of scalar invariants,
which are local quantities that are advected by the flow. 
The hierarchy is generated by a recursion operator \eqref{inv.recursionop}
applied to the basic invariant $S$. 
In the case of entropic \esos/, 
radial fluid flow possesses two additional hierarchies of scalar invariants \eqref{J1.hierarchy} and \eqref{J2.hierarchy}.
These yield corresponding integral invariants \eqref{integral.inv.J1J2hierarchy},
which describe more advected conserved integrals. 
It will be interesting to connect these results to invariants defined on 
the characteristic curves of the fluid flow equations using the methods in \Ref{KapZab,Kap}. 

A computational classification of non-trivial conserved integrals given by a first-order conserved density \eqref{1storder.consdens.quadr.Ur} 
also has been carried out. 
The classification shows that such conserved integrals exist only for entropic \esos/ 
and are equivalent to particular integral invariants \eqref{newdens1} and \eqref{newdens2} (modulo trivial conserved integrals). 

An investigation of symmetries, Hamiltonian structure, and Casimirs 
related to all of the conserved integrals has been completed recently \cite{AncSeiWol}. 

Finally, two applications of the lowest-order new conserved integrals 
have been outlined (cf Section~\ref{sec:scaling})
for the analysis of smooth flows, shocks, and self-similar flows. 
It will be fruitful to explore these applications in detail and to see if 
the higher-order new conserved integrals in the two hierarchies may have similar
applications. 
In particular, 
imploding self-similar flows have been obtained very recently in \Ref{MerRapRodSze} 
for barotropic \esos/ with a general power law form (cf cases (B) and (C) in Section~\ref{sec:eqns});
it will be interesting to study self-similar non-isentropic flows 
with the \esos/ shown in Table~\ref{table:scalingsymms}. 

All of the preceding results carry over to radial gas dynamics through the well-known 
equivalence between the respective governing equations of 
$n$-dimensional gas dynamics and $n$-dimensional compressible fluid flow. 
Specifically, 
when a conserved integral or an invariant is expressed solely in terms of $t$, $r$, $U$, $\rho$, $p$, $e$, 
then it manifestly holds for both radial fluid flow and radial gas dynamics. 

In a different direction, 
recent work \cite{AncWeb2020} 
on nonlocal vorticity invariants for compressible non-radial fluid flow in three dimensions
has found new nonlocal conservation laws
involving advective potentials, derived in terms of various thermodynamic quantities. 
This method can be applied to radial fluid flow in $n$ dimensions. 
As an example, 
consider a scalar potential $\mu$ defined in terms of the temperature $T$ by 
\begin{equation}
\frac{d\mu}{dt} = T . 
\end{equation}
A nonlocal conservation law is now given by 
\begin{equation}
\Phi^t = U -\mu S_r,
\quad
\Phi^r = \tfrac{1}{2}U^2 +e + p/\rho  -\mu U S_r
\end{equation}
which has non-zero moving flux $\Psi = e + p/\rho  -\tfrac{1}{2}U^2$. 
The resulting conserved integral generalizes the enthalpy-flux \eqref{enthaplyflux.integral} 
to hold for a general \eos/. 
Hence a new non-advected quantity is obtained. 
A full exploration of nonlocal conservation laws and nonlocal invariants 
for radial fluid flow will be left for future work.

\appendix

\section{Computational remarks for scalar invariants}\label{app:invs.computation}

The first step is substitution of $J(t,r,U,\rho,S,U_r,\rho_r,S_r)$ 
into the determining equation \eqref{scalarinv.deteqn}, 
followed by elimination of $t$-derivatives through use of 
the radial Euler equations \eqref{radialU.eqn}--\eqref{radialS.eqn}. 
This yields a single PDE which splits with respect to $U_{rr}$, $\rho_{rr}$, $S_{rr}$, 
giving an overdetermined system of 9 PDEs. 
The unknowns are $J$, $p(\rho,S)$, and $n$, which are subject to the conditions
$p\neq$const.\ and $n\neq1$. 
These unknowns appear nonlinearly in the system, and hence the problem is nonlinear. 

The Maple command `rifsimp' is used to obtain a complete classification of 
all cases in which the system can be brought into a consistent involutive form. 
This yields two cases, distinguished by a general \eos/ for $p$,
and an entropic \eos/ for $p$. 
In each case, the resulting simplified system is first-order and linear in $J$,
and thus it can be easily solved. 
The case tree of solutions leads to the results in Theorem~\ref{thm:1storder.invs}
for scalar invariants $J(t,r,U,\rho,S,U_r,\rho_r,S_r)$. 

The same method can be used to find $J$ with dependence on higher $r$-derivatives of $U$, $\rho$, $S$, 
up to any specified (finite) differential order.

\section{Computational remarks for conserved densities}\label{app:consdens.computation}

\subsection{Kinematic case}
The steps for classifying kinematic conserved densities \eqref{kinematic} 
start from 
\begin{equation}
(D_t \Phi^t)|_\Esp = 
\Phi^t_t 
-\Phi^t_U\big( U U_r + (p_S S_r + p_{\rho} \rho_r)/\rho \big)
-\Phi^t_\rho\big( (U\rho)_r +(n-1) U \rho/r \big)
- \Phi^t_S\big( U S_r \big)
\label{DtPhi}
\end{equation}
Here $E_v = \sum_{i\geq0}(-D_r)^i \partial_{\partial_r^i v}$ 
denotes the radial Euler operator with respect to a variable $v$; 
this operator coincides with the variational derivative $\delta/\delta v$. 
Substitution of expression \eqref{DtPhi} into the determining equations \eqref{deteqns}, 
followed by splitting each equation with respect to $r$-derivatives of $U$, $\rho$, $S$,
yields an overdetermined system of  14 PDEs, 
with $\Phi^t$, $p(\rho,S)\neq$const.\ and $n\neq1$ being the unknowns. 
Since the unknowns appear nonlinearly in the system, the problem is nonlinear. 

This system can, in principle, be solved by applying the same steps outlined for solving the overdetermined system for scalar invariants. 
However, `rifsimp' is unable to return all cases in the classification, due to their complexity. 
This computational difficulty can be by-passed by first dividing the classification into 
the following distinct (non-overlapping) cases for the \eos/: 

(a) general $p=f(\rho,S)$, $f_\rho\not\equiv0$, $f_S\not\equiv0$;

(b) entropic $p=\k(S)$, $\k\neq$const.;

(c) polytropic $p=\k \rho^{q}$, $q\neq0$ ;

(d) barotropic $p=f(\rho)$, $p\neq\k \rho^{q}$.

In each case, `rifsimp' is able to return a complete classification of 
all subcases in which the system can be brought into an involutive form. 
The resulting systems are second-order and linear in $\Phi^t$. 
They can be solved by the following integration steps. 

First, solve the PDE(s) that involve only $p$, 
and substitute the solution for $p$ back into the system. 
Second, solve simplest PDEs for $\Phi^t$ (e.g. one-term or two-term equations); 
the solution will involve some arbitrary functions/constants. 
Next, substitute back into the system; 
if at any step there is a variable that does not appear in all arbitrary functions, 
then split the system with respect to that variable. 
Solve simplest PDEs involving a single function,
and then eliminate any redundant functions/constants after the solution has been substituted into $\Phi^t$. 
Simplify the pruned system by `rifsimp', and repeat the previous steps 
until all equations have been solved. 
Finally, split $\Phi^t$ with respect to any free constants/functions. 

These basic steps work for cases (a), (b), (c). 
In case (d), an additional step of changing variables is needed 
after one of the integrations; the original variables are substituted back in the final step.

The last step consists of deriving $\Phi^r$ by inverting the total $r$-derivative
in the conservation law equation \eqref{Phi.eqn}. 
The inversion can be done by a straightforward integration by parts process, 
starting from terms of highest-order derivatives in the righthand side 
and descending to terms of first-order derivatives. 

This leads to the results in Theorem~\ref{thm:kinematic.consintegrals}
for kinematic conserved integrals. 

\subsection{First-order quadratic case}
A more refined solving process is needed 
for solving the determining equations \eqref{deteqns} 
for conserved densities $\Phi^t$ with the first-order form \eqref{1storder.consdens.quadr.Ur},
since there are solutions such that $\Phi^t=r^{1-n}D_r\Theta$ is a trivial density,
where $\Theta(t,r,U,\rho,S)$ is an arbitrary differentiable function. 

These trivial solutions can be eliminated by imposing the condition that 
at least one of the three expressions 
$E_U(r^{n-1}\Phi^t)$, $E_\rho(r^{n-1}\Phi^t)$, $E_S(r^{n-1}\Phi^t)$ 
is not identically zero. 
However, even under these conditions, there are solutions given by the sum of 
a kinematic conserved density and an arbitrary trivial solution. 
For such solutions, 
the expressions $E_U(r^{n-1}\Phi^t)$, $E_\rho(r^{n-1}\Phi^t)$, $E_S(r^{n-1}\Phi^t)$ 
will be zeroth order. 
Hence, all unwanted solutions can be eliminated through the condition that 
at least one of $E_U(r^{n-1}\Phi^t)$, $E_\rho(r^{n-1}\Phi^t)$, $E_S(r^{n-1}\Phi^t)$ 
has non-zero differential order:
\begin{equation}\label{ineqn}
\sum_{v=U,\rho,S} \big(\partial_{(U_r,\rho_r,S_r,U_{rr},\rho_{rr},S_{rr})} E_v(r^{n-1}\Phi^t)\big)^2 
\not\equiv 0
\end{equation}
where $\partial_{(U_r,\rho_r,S_r,U_{rr},\rho_{rr},S_{rr})}$ 
denotes the set of partial derivatives with respect to the indicated variables. 

To proceed, 
the determining equations are split with respect to the $r$-derivatives of $U$, $\rho$, $S$ 
that do not appear in the unknowns $\phi_0$, $\phi_1$, $\phi_2$. 
This yields an overdetermined system of 96 PDEs, 
plus the inequation \eqref{ineqn}, 
with the additional unknowns $p(\rho,S)\neq$const.\ and $n\neq1$. 

The three cases (a), (b), (c) can be handled in a combined way, 
with the condition $p_\rho\not\equiv 0$. 
In this combined case, `rifsimp' finds that the system is inconsistent, 
and hence no solutions describing non-trivial first-order conserved densities exist. 

In the remaining case (d), `rfisimp' returns a consistent system. 
The basic steps outlined for the kinematic case
enable the integration of most of this system, 
until a set of 6 coupled PDEs remain which are not simple to integrate. 
The solution is obtained through an intricate combination of 
changes of independent and dependent variables, direct integrations, 
and simplifications that ultimately reduce the PDEs into a triangular, first-order system 
which can be integrated by standard techniques. 
A crucial part of this process is the elimination of redundant functions in the $\phi$'s
as well as the use of integration by parts to remove trivial terms in $\Phi^t$. 
Additionally, manual case splitting is invoked twice, 
which aids in allowing some of the steps to work. 

This leads to the final result summarized in Theorem~\ref{thm:1storder.consintegrals}.

\section*{Acknowledgements}
SCA is supported by an NSERC Discovery research grant.

On behalf of all authors, the corresponding author states that there is no conflict of interest. 
Data sharing not applicable to this article as no datasets were generated or analysed during the current study.

\end{document}